# PUBLISH-SUBSCRIBE FRAMEWORK FOR EVENT MANAGEMENT IN IOT-BASED APPLICATIONS

Nguyen Dinh Trung Truc, Nguyen Mai Bao Quan, and Pham Hoang Anh
Ho Chi Minh City University of Technology, Viet Nam

Email: {51304505, 51303225, anhpham}@hcmut.edu.vn

**ABSTRACT**

The incredible growth of sensors and microcontroller units makes the task of real-time event monitoring in the Internet of Things (IoT) based applications easier and more practical. In order to effectively support the event management in IoT-based applications, we propose a framework that is based on the publish-subscribe model for detecting events from IoT sensor nodes and sending notifications to subscribers (end-users) via Internet, SMS, and Calling. With the exception of the advantages inherited from the publish-subscribe model, the further advantages of the proposed framework are the ease of use in terms of user configuration without any need of technical skills; the aid of security mechanisms to prevent network intrusion; and the minimum hardware resource requirement. Additionally, the proposed framework is applicable and adaptable to various platforms since it has been developed by using Boost C++ Libraries and CMake. To evaluate the proposed framework, we develop a prototype of a real-time event monitoring system that is also presented in this paper.

## 1. INTRODUCTION

In the Internet of Things (IoT) paradigm, pervasive presence around us of a variety of things or objects which, through unique addressing schemes, are able to interact with each other and cooperate with their neighbors to reach common goals (Atzori, Iera, & Morabito, 2010). A noticeable application of this technology is the development of remote monitoring systems in which data are collected from various sensors. For example, people will be able to observe any event occurred inside their houses such as power outage, and intrusion or sudden change of temperature even when they are outside. There have been many implementations of such system in recent studies (Marimuthu CN, 2015; Tserng, 2013). In this paper, we contribute to this field of study by proposing a publish-subscribe framework for event management including event detection and notification for IoT-based applications.

Publish-subscribe is a popular paradigm for users to express their interests ("subscriptions") in certain kinds of events ("publications") (Demers, Gehrke, Hong, Riedewald, & White, 2006) and are subsequently notified of any event, generated by a publisher, which matches their registered interest (Eugster, Felber, Guerraoui, & Kermarrec, 2003). The publisher does not send messages directly to subscribers, however, it characterizes the published messages into classes without the need of any information about subscribers. To accomplish this task, an intermediary, called a "message broker" or "event bus" as described in Fig. 1, receives published messages, and then forwards them to the corresponding subscribers. A big challenge of a publish-subscribe system is to timely deliver the published events to all interested subscribers. Such system provides greater network scalability than traditional client-server.

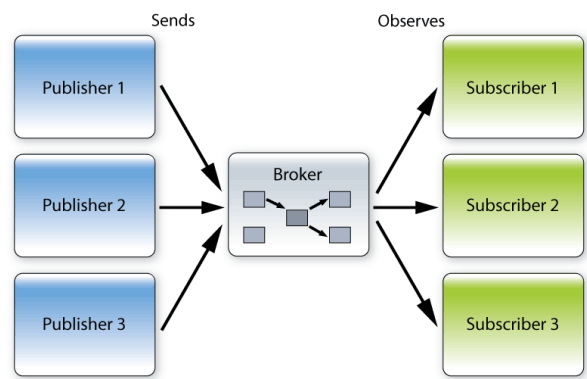

Fig 1. A publish-subscribe model

The rest of this paper is organized in the following manners. Section 2 describes the overall architecture of the proposed framework. Then, Section 3 presents a software-based design for the Gateway that is a major component in the proposed framework. Section 4 presents our prototype and the experiments to evaluate the proposed framework. Finally, the concluding remarks are given in Section 5.

## 2. THE PROPOSED FRAMEWORK

The proposed framework is designed for event management in IoT-based applications in which each event is characterized by a specific condition. Table 1 shows an example for characterizing an event by a data structure called Channel that consists of a channel ID and the condition of the corresponding event.

Table 1. An example of event characterization

| Channel ID | Condition |
|---|---|
| 0 | Power outage |
| 1 | Temperature is higher than 50 °C |
| 2 | Light level is below 20 lux |
| … | … |

### 2.1 Overall Architecture

Fig. 2 describes the overall architecture of the proposed framework that consists of three major components such as the Sensor node, Client, and Gateway.

- The Sensor node collects data from sensors and transfers those data to the Gateway.
- The Client represents the end-user, each Client is identified by a phone number. It becomes a subscriber when it subscribes to a specific event of the system.
- The Gateway acts as a broker in the publish-subscribe model. It receives information from the publishers (i.e. sensor nodes) and the information is filtered into predefined channels. Then, for each channel, the Gateway will notify to all corresponding subscribers when the received value satisfies a predefined condition.

Fig. 3 describes the association between three components in the proposed framework. A client may associate with many Gateways and a Gateway may have many clients. A Sensor node can associate with only one Gateway meanwhile a Gateway may have many Sensor nodes.

Fig. 4 describes the network architecture of the proposed framework in which the Gateway acts as a Wi-Fi access point so that it can connect to all Sensor nodes.

The Gateway is also connected to a router which has direct access to the Internet. Users can configure the system via Bluetooth and Internet, receive notifications via Internet and Global System for Mobile communication (GSM) including Short Message Service (SMS) and Calling. All the messages exchanged between these components are formatted with JavaScript Object Notation (JSON) (Severance, 2012).

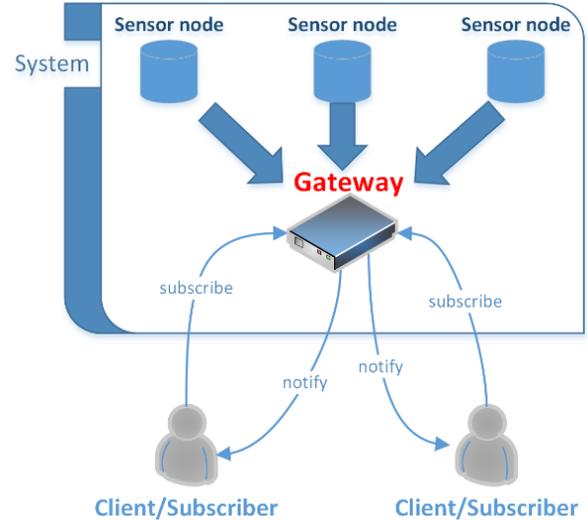

Fig 2. Overall architecture

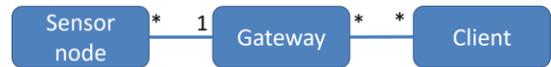

Fig 3. Association of Sensor node, Gateway and Client

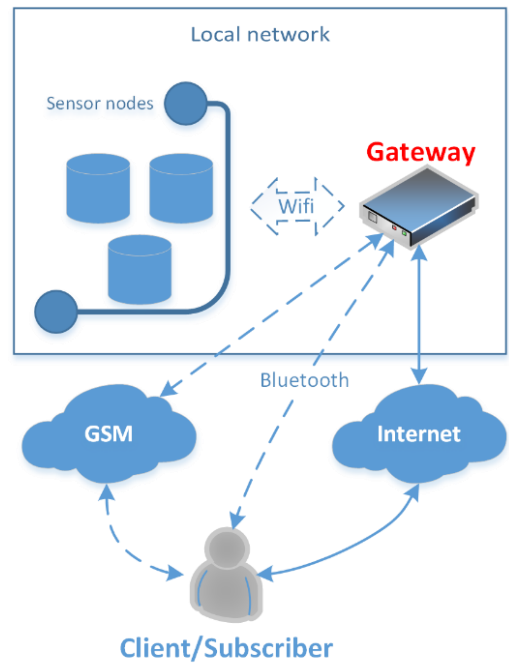

Fig 4. Network architecture

*2.1.1 Sensor nodes and Gateway Communication*

The Sensor nodes and the Gateway form a local area network (LAN). When a Sensor node connects to the Wi-Fi network of the Gateway, it will be dynamically assigned an IP address. The IP address of the Gateway is static to simplify the configuration on Sensor nodes. The TCP will be used to utilize the reliable data transfer mechanism.

*2.1.2 Gateway and Clients Communication*

Connection via the Internet enables users to configure the system and receive notifications. For user configuration, a Client directly connects to the Gateway using TCP and sends commands. For receiving notifications, we use Firebase Cloud Messaging (FCM) provided by Google ("Firebase Cloud Messaging", 2014). It is a cross-platform messaging solution that provides the capability of reliably delivering messages to a Client application. Each Client application is associated with a unique registration token (FCM ID) which is later used by the Gateway to address the messages. Because the connection is over the Internet, security is an issue

The second choice in communication is to send notifications by SMS. This solution ensures the message will be arrived to the Subscribers when they are within a GSM coverage area. On the contrary, using SMS results in the increase of cost for each event notification. Additionally, some emergency events can be configured to notify by calling the Subscribers.

Bluetooth is an another choice to configure the system when users are within the Bluetooth range which is typically less than 10 meters, up to 100 meters ("Bluetooth Range: 100m, 1km, or 10km?", 2015). The advantage of this method is to prevent the network intrusion. In the proposed framework, RFCOMM protocol is adopted in order to provide approximately the same service and reliability guarantees as TCP (Huang & Rudolph, 2007).

**2.2 Security Issue**

In order to secure the connection between the Clients and the Gateway via Internet as mentioned above, we propose two conventional security mechanisms including (1) User-based authorization and (2) Secure Sockets Layer (SSL) (Shinozaki & Arai, 2014) security technology.

*2.2.1 User-based authorization*

There are two types of account in the proposed framework including the Subscriber and Admin.
- Subscriber: this account is authorized to subscribe to events needed to be kept track of and receive notifications from the system.
- Admin: the users obtaining the Admin privileges are able to have a full control of the system such as editing Subscribers, updating software, and changing account password.

Fig. 5 describes the association between these two accounts and the system. The Subscriber and Admin may associate with many Gateways meanwhile a Gateway may have many Subscribers but just one Admin.

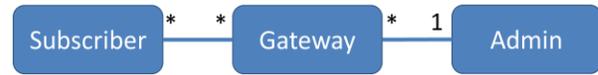

Fig 5. Account association

*2.2.2 SSL security technology*

SSL is currently one of the most common and efficient method to establish an encrypted link. SSL ensures that all data passed thru the connections between the network entities will remain private and integral. In the proposed framework, we utilize an open source project OpenSSL (Park, Han, & Lee, 2014) that provides a robust, commercial-grade, and full-featured toolkit for the TLS/SSL protocols.

**2.3 Operating system**

Nowadays, there are plenty of operating systems (OS) specialized to be used in embedded system such as Windows CE, Windows 10 IoT Core and some Linux distributions including OpenWRT, Raspbian and Android. Therefore, we have to deliver a strategy toward an appropriate OS for the proposed framework which consists of the following characteristics:
- Responsiveness: a real-time OS is needed because the event detection and notification must be deterministic in terms of time.
- Networking and Connectivity: it is essential to have an OS that supports various connectivity technologies such as Serial, Ethernet, Wi-Fi and Bluetooth which enables the Gateway to connect to the Sensor nodes as well as the Clients flexibly.
- Security: the OS has to support OpenSSL to secure the connection between the Gateway and the Clients via Internet.
- Portability: the OS should be able to operate on various hardware platforms in order to reduce the rework effort due to some changes in the hardware.
- Cost: an open source OS is preferred to minimize the product cost.

In the proposed framework, we create a Linux distribution based on the Yocto Project ("Yocto Project", 2010) in order to satisfy all the characteristics as abovementioned. Normally, the standard Linux kernel comes as a soft real-time OS. However, the proposed framework would require a better scheduling scheme. Therefore, a real-time preemption patch for Linux (Clark, 2013) will be adopted.

Additionally, in order to achieve the adaptability to various operating systems, Boost C++ Libraries and CMake are used. Boost ("Boost C++ Libraries", 2005) is a set of C++ libraries that provides support for many tasks and structures such as multithreading, object serializing and file system manipulation. The key advantage is that it provides some OS abstraction libraries to manage threads and file system so that the framework is able to be compiled on different platforms. CMake (Clemencic & Mato, 2012) controls the build process of a software using a compiler-independent method and generates native makefiles that can be used in a specific environment. The obvious advantage of CMake is that we only need to create only one set of build scripts for various platforms.

### 2.4 User configuration

Each connection between the Client and the Gateway via Bluetooth or the Internet is referred as a session. Each session will be structured in a sequence diagram as illustrated in Fig. 6.

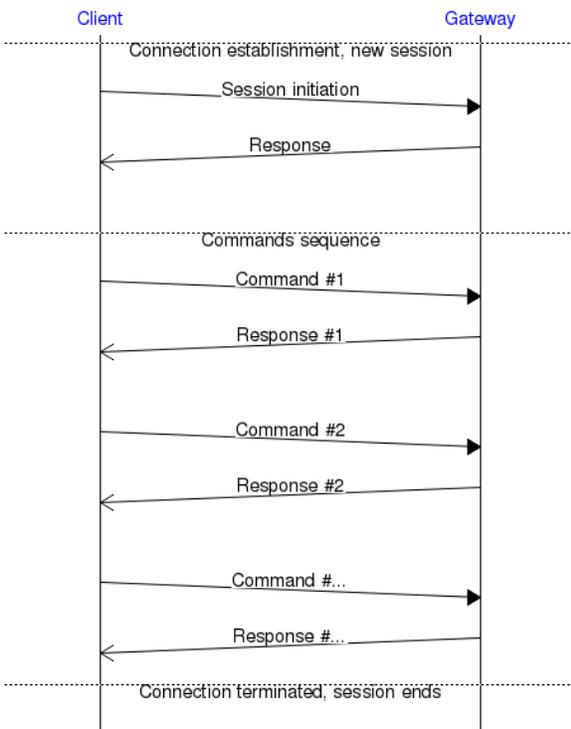

Fig 6. Sequence diagram of a session

A session consists of three steps including (1) connection establishment, (2) commands sequence, and (3) connection terminated.

- *Connection establishment*: the Client sends a message indicating the initiation of a new session with an account and the corresponding password. Then, the Gateway responses the result of the session initiation. If the session is successfully established, it will come to the Commands sequence step, or else, come to the Connection terminated step.
- *Commands sequence*: the Client sends some predefined user configuration commands and for each command sent, it waits for the response from the Gateway. Since the proposed framework supports two types of account and each account may have each own set of commands, the Gateway must determine the validity of the received message.
- *Connection terminated*: the session ends when the connection is terminated, it occurs when either side closes the socket on its own or the connection between them is unexpectedly disconnected.

## 3. A SOFTWARE-BASED DESIGN FOR THE GATEWAY

The Gateway is a key component in the proposed framework. Almost all functions and characteristics of the proposed framework will be implemented on the Gateway. In this section, we present a software-based design in object-oriented for the Gateway as depicted in Fig. 7.

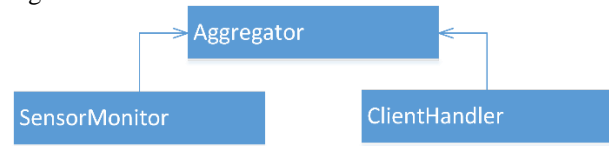

Fig 7. Software-based design overview for the Gateway

### 3.1 Aggregator

Aggregator is a class that stores all the data such as accounts' password and some data structures including the value of each channel, a mapping between a channel and a list of its Subscribers' address. The address consists of a phone number and a FCM ID. Additionally, it provides methods to manage those data.

### 3.2 SensorMonitor

SensorMonitor is responsible for receiving data from Sensor nodes and determining the occurrence of an event to notify the Subscribers. It monitors all the sockets established between the Gateway and the Sensor nodes. For each received packet, it parses the JSON object and gets the value as well as the channel ID, then it checks the channel's condition to determine whether the event occurs. If the condition is satisfied, SensorMonitor will create a new thread to notify the channel's Subscribers.

### 3.3 ClientHandler

ClientHandler is a class that handles the session created with the Client and provides methods to execute commands in user configuration. It parses the received command from the Client and executes the corresponding method. Since the Gateway supports multiple ways to communicate with Clients, the communication functions are virtual methods. Moreover, for each type of connection, we create a particular class that inherits from the class ClientHandler and implements the corresponding communication methods such as ClientBLE for Bluetooth, ClientGSM for GSM, and ClientTCP for TCP. The relationship between these classes is described in Fig. 8.

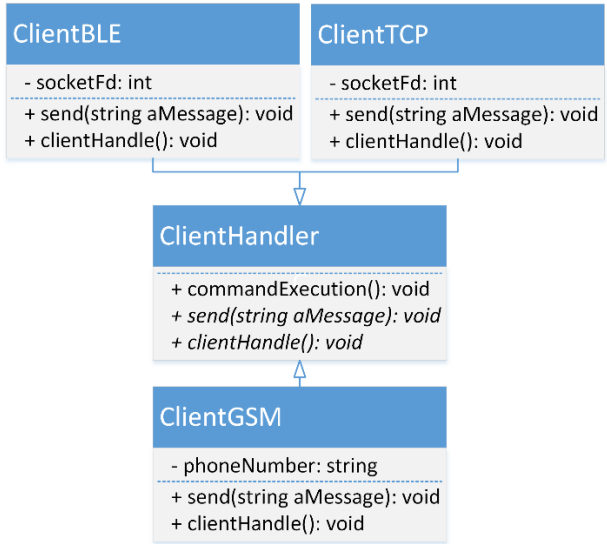

Fig 8. Inheritance of ClientHandler

## 4. PROTOTYPE AND EVALUATION

To evaluate the proposed framework, we implement a prototype that is called Power-Signal Detection Unit (PSDU) for detecting AC power outage and sending notifications to the end-users via SMS and the Internet. The design of the PSDU is described in Fig. 9.

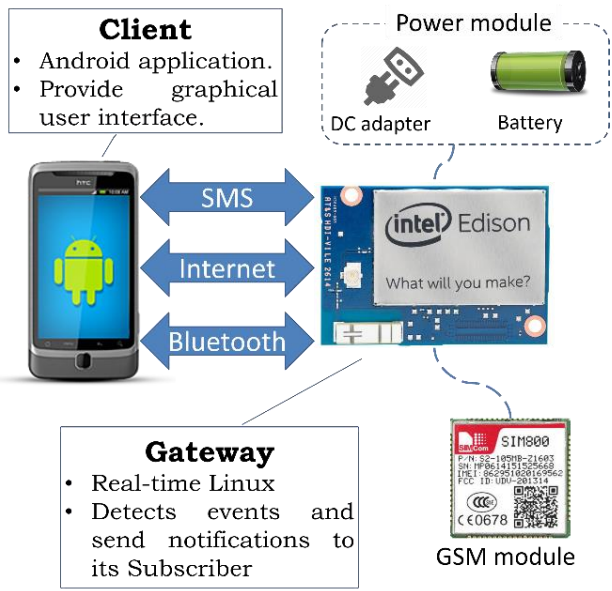

Fig 9. Power-signal detection unit (PSDU)

### 4.1 Hardware

The Intel® Edison Module that is a tiny, SD-card-sized computing chip designed for building Internet of Things (IoT) and wearable computing products (Balani, 2016), will be used as the computing unit in our prototype. The Edison module contains a high-speed, dual-core processing unit, integrated Wi-Fi, Bluetooth low energy, storage and memory, and a broad spectrum of input/output (I/O) options for interfacing with user systems. The module consists of an Intel® Atom™ processor operating at a clock speed of 500 MHz, 1GB of RAM and 4 GB of managed flash memory. By default, the Yocto Linux operating system is installed in flash memory. The design of the Gateway will be implemented in this module.

Since the Intel Edison has already supported Bluetooth and Wi-Fi, we further use SIM800, a GSM module for supporting SMS and calling notification. The SIM800 module supports Quad-band 850/900/1800/1900MHz, it can transmit Voice, SMS and data information with low power consumption (SIMCom, n.d.). This module communicates with the Intel Edison via the serial port using a set of AT commands which was initially implemented by the manufacture.

Additionally, another important component of this PSDU prototype is the power module that is a backup battery to power the system when the main power supply is off. Moreover, this system has to be able to charge the battery when the main supply is available. Therefore, a circuit based on (Chu, 2008) is implemented as a Lithium battery charger with load sharing. The schematic of this circuit is shown in Fig. 10.

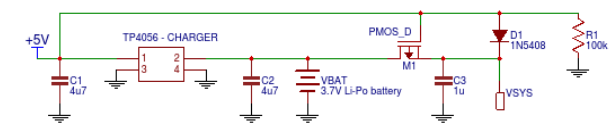

Fig 10. Load sharing battery charger

### 4.2 JSON-based commands

In this prototype, a set of JSON-based commands for the communication between the Client and the Gateway is defined and each command is identified by the tag "action" as represented in Table 2.

The Gateway responses with a JSON object which has to include two tags "result" and "desc" in order to indicate the validity of the command and further description. Moreover, depends on the characteristic of the command, the response may have some more tags such as a list of current subscribers. All of the information about the communication is summarized in a separate protocol document.

Table 2. Set of main commands

| Command | "action" | Description |
|---|---|---|
| Session Initiation | "SessionInitiation" | Initiates a new session. Requires a specific account name and password. Available for Admin and Subscriber. |
| Change account password | "ChangePassword" | Change the password of an account. Requires an account name and the new password. Available for Admin only. |
| Get subscriber list | "GetSubscriberList" | Get the list of current Subscriber of a specific event. Requires an event name. Available for Admin only |
| Delete a subscriber | "DelSubscriber" | Delete a subscriber from a specific event. Requires the subscriber's phone number and the event name. Available for Admin only. |
| Add subscriber | "AddSubscriber" | Add a subscriber to an event. Requires the subscriber's phone number, FCM ID and the event name. Available for Admin only. |
| Subscribe | "Subscribe" | Subscribe to an event. Requires the phone number, FCM ID and the event name. Available for Subscriber only. |
| Unsubscribe | "Unsubscribe" | Remove a subscription. Requires the phone number and the event name. Available for Subscriber only. |
| Update status | "Update" | Query the current status of an event. Available for Admin and Subscriber. |

**4.3 Experiments and Evaluation**

All major components of the product have been assembled as shown in Fig. 11.

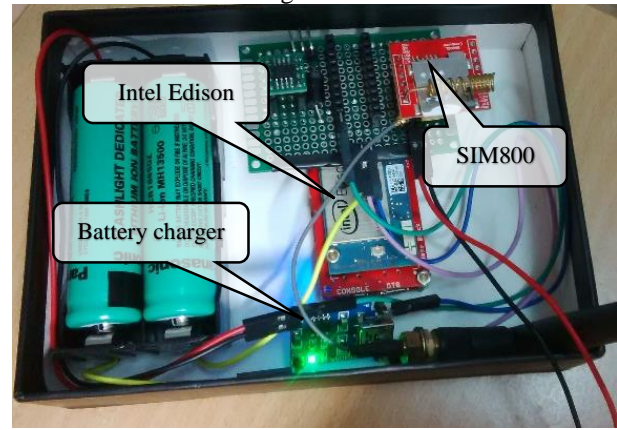

Fig 11. Prototype of the proposed framework

A screenshot for the command Add subscriber in Android-based user application is shown in Fig. 12. The mobile application interface enables users to easily configure the system without any need of technical skills.

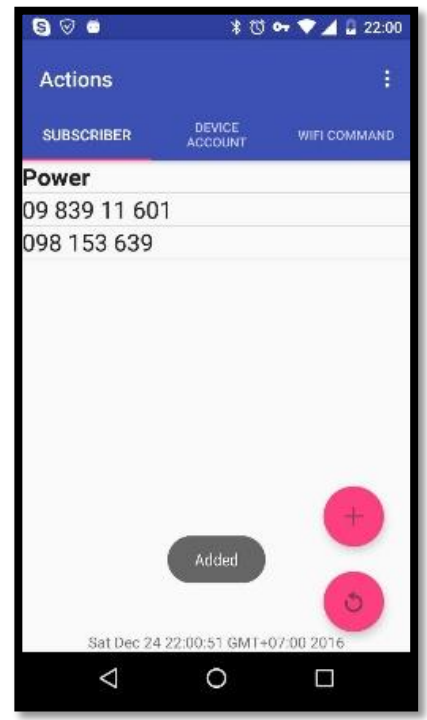

Fig 12. Screenshot of the mobile application

In the first experiment, we calculate the responsiveness of the system to an event. In this case, the event is power outage. We use **boost::posix_time** to estimate the time duration from the occurrence of the event to the delivery of the first SMS (wall clock time). The testing procedure is that we unplug the system from the main power supply and get the returned elapsed time. In this experiment, we carry out the procedure for ten times and the average elapsed time is **4262.5 milliseconds**.

For the second experiment, we calculate the memory and CPU usage of the Gateway by using Linux's "top" command. The memory usage depends on the number of Subscribers, thus, we increase the number of Subscriber by sending the Subscribe command. The results are summarized in Table 3. The CPU usage of the program is less than 1% in normal condition because threads are in the sleep state most of the time.

Table 3. Memory usage

| Number of Subscribers | Memory usage (KB) (Resident Set Size) |
|---|---|
| 0 | 2688 |
| 30 | 2708 |
| 100 | 4900 |
| 1000 | 4984 |
| 2000 | 5480 |
| 5000 | 6816 |

**5. CONCLUSIONS**

In this paper, we propose a framework based on the publish-subscribe model for event management in IoT-based applications. We also develop a prototype to evaluate the proposed framework. The experimental results initially show that the framework achieves minimum hardware resources and the ease of use in terms of user configuration. Additionally, the proposed framework is applicable and adaptable to various platforms by using Boost C++ Libraries and CMake. Therefore, it helps reducing effort in developing a real-time monitoring system for the IoT-based applications.

**REFERENCES**

Atzori, L., Iera, A., and Morabito, G., The internet of things: A survey, *Computer networks*, vol. 54, no. 15, pp. 2787-2805, 2010.

Demers, A., Gehrke, J., Hong, M., Riedewald, M., and White, W., Towards expressive publish/subscribe systems, *International Conference On Extending Database Technology*, pp. 627-644, 2006.

Tserng, H., Application of Wireless Sensor Network to the Scour Monitoring System of Remote Bridges, *International Journal Of Engineering And Technology*, 641-647, 2013

Marimuthu CN, J., Wearable Real Time Health and Security Monitoring Scheme for Coal Mine Workers, *Journal Of Electrical & Electronic Systems*, vol. 4, no. 2, 2015.

Eugster, P., Felber, P., Guerraoui, R., and Kermarrec, A., The many faces of publish/subscribe, *ACM Computing Surveys*, vol. 35, no. 2, pp. 114-131, 2003.

Severance, C., Discovering JavaScript Object Notation, *Computer*, vol. 45, no. 4, pp. 6-8, 2012.

Ford, B., Srisuresh, P., & Kegel, D. (2005, April). Peer-to-Peer Communication Across Network Address Translators. In USENIX Annual Technical Conference, General Track (pp. 179-192).

*Firebase Cloud Messaging*, *Firebase*, Retrieved 5 December 2016, from https://firebase.google.com/docs/cloud-messaging/, 2014.

*Bluetooth Range: 100m, 1km, or 10km?*, *Bluair.pl*, Retrieved 7 November 2016, from http://www.bluair.pl/bluetooth-range, 2015.

Shinozaki, J. and Arai, M., Secure Socket Layer Visualization Tool with Packet Capturing Function, *International Journal Of Future Computer And Communication*, vol. 3, no. 3, pp. 187-190, 2014.

Park, G., Han, H., and Lee, J., Design and Implementation of Lightweight Encryption Algorithm on OpenSSL, *The Journal Of Korean Institute Of Communications And Information Sciences*, vol. 39B, no. 12, pp. 822-830, 2014.

*Yocto Project*, *yoctoproject.org*, Retrieved 25 June 2016, from https://www.yoctoproject.org, 2010.

Clark, L., *Intro to Real-Time Linux for Embedded Developers*, *Linux.com | The source for Linux information*, Retrieved 15 August 2016, from https://www.linux.com/blog/intro-real-time-linux-embedded-developers, 2013.

*Boost C++ Libraries*, *Boost.org*, Retrieved 25 December 2016, from http://www.boost.org/, 2005.

Clemencic, M. and Mato, P., A CMake-based build and configuration framework, *Journal Of Physics: Conference Series*, vol. 396, no. 5, 052021, 2012.

Balani, N., *What Is the Intel® Edison Module? | Intel® Software*, *Software.intel.com*, Retrieved 15 August 2016, from https://software.intel.com/en-us/articles/what-is-the-intel-edison-module, 2016.

SIMCom, *SIM800 - GSM/GPRS Module,* Retrieved 22 July 2016, from http://simcomm2m.com/En/module/detail.aspx?id=138.

Chu, B., Designing A Li-Ion Battery Charger and Load Sharing System With Microchip's Stand-Alone Li-Ion Battery Charge Management Controller, © *2008 Microchip Technology Inc,* Retrieved from http://ww1.microchip.com/downloads/en/AppNotes/01149c.pdf, 2008.

**PHOTOS AND INFORMATION**

| | |
|---|---|
| 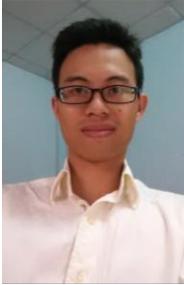 | **Nguyen Dinh Trung Truc** is a senior student at the Faculty of Computer Science and Engineering, Ho Chi Minh City University of Technology (HCMUT). His current interests include Internet of Things, Embedded System, Computer Architecture, Operating System, Data Structures and Algorithms. |
| 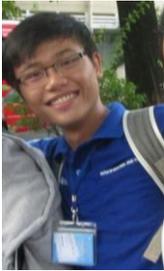 | **Nguyen Mai Bao Quan** is a senior student at the Faculty of Computer Science and Engineering, Ho Chi Minh City University of Technology (HCMUT). His current interests include Database, Mobile Application Development, Operating System, Data Structures and Algorithms. |
| 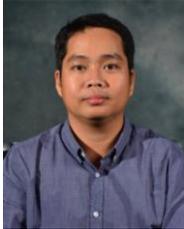 | **Pham Hoang Anh** received his BEng in Computer Science and Engineering in 2005 from Ho Chi Minh City University of Technology (HCMUT), Vietnam. From 2005 to 2008, he was an assistant lecturer at the Faculty of Computer Science and Engineering, (CSE-HCMUT). In 2010 and 2014, he received his MSc and PhD in Communications Engineering from MYONGJI University, Korea, respectively. He is currently the Head of the IoT-Lab, senior researcher, and lecturer at CSE-HCMUT. His research interests include fault-tolerant networks, IoT, big data, and data communications. |